# Diagnosis of COVID-19 Cases from Chest X-ray Images Using Deep Neural Network and LightGBM

Mobina Ezzoddin
*Electrical and Computer Engineering Department*
*Semnan University*
Semnan, Iran
mobina.ezzoddin@gmail.com

Hamid Nasiri
*Computer Engineering Department*
*Amirkabir University of Technology*
Tehran, Iran
h.nasiri@aut.ac.ir

Morteza Dorrigiv
*Electrical and Computer Engineering Department*
*Semnan University*
Semnan, Iran
dorrigiv@semnan.ac.ir

*Abstract—* **The Coronavirus was detected in Wuhan, China in late 2019 and then led to a pandemic with a rapid worldwide outbreak. The number of infected people has been swiftly increasing since then. Therefore, in this study, an attempt was made to propose a new and efficient method for automatic diagnosis of Corona disease from X-ray images using Deep Neural Networks (DNNs). In the proposed method, the DensNet169 was used to extract the features of the patients' Chest X-Ray (CXR) images. The extracted features were given to a feature selection algorithm (i.e., ANOVA) to select a number of them. Finally, the selected features were classified by LightGBM algorithm. The proposed approach was evaluated on the ChestX-ray8 dataset and reached 99.20% and 94.22% accuracies in the two-class (i.e., COVID-19 and No-findings) and multi-class (i.e., COVID-19, Pneumonia, and No-findings) classification problems, respectively.**

## I. INTRODUCTION

In late December 2019 and at the beginning of the 2020, a new type of acute respiratory disease was identified in Wuhan, Hubei Province, China for the first time [1]. The disease spread rapidly, while the number of people infected with it is still increasing swiftly and causing death of a large number of them [2]. The disease was caused by Severe Acute Respiratory Syndrome Coronavirus 2 (SARS-CoV-2) and called COVID-19 virus by the World Health Organization [3]. The symptoms of this disease, which might be mild or severe, included fever, dry cough, sore throat, headache, fatigue, and shortness of breath [4], [5]. The best ways to prevent the spread of this virus have been to observe a proper physical distance at all times, avoid touching the eyes, nose, and mouth with unwashed hands, and use a mask to completely cover the nose and mouth [6].

Generally, there are several tests for the diagnosis of Coronavirus disease, including the Real-Time Reverse Transcription-Polymerase Chain Reaction (RT-PCR) test. This test depends on specialized equipment and is time-consuming. It does not have high accuracy and the results may include false-negative [7], [8]. Other tests have incorporated radiographic imaging of the chest, such as Computed Tomography-Scan (CT-Scan) and Chest X-Ray (CXR) [7], [9]. The equipment required for these types of tests are lightweight and portable, while their imaging takes about 15 seconds for each patient [10]. Radiographic imaging devices are available in most specialized laboratory centers, clinics, and hospitals. Analysis and pre-examination of CXR images taken from the patients with COVID-19 showed that this type of test was a fast approach and one of the most cost-effective methods for diagnosing the mentioned patients. Therefore, this type of image was applied in this study [11].

Deep Learning (DL) is a subset of a broader family of machine learning algorithms based on artificial neural networks. A Convolutional Neural Network (CNN) is a type of neural network architecture, which is utilized as an efficient and intelligent architecture in pattern recognition and image classification, especially in the field of medical and radiological images and disease diagnosis [12]–[14].

CNNs are multilayered neural networks, through which images can be properly recognized by using convolution and pooling layers to reduce their main features [15]. There are varied approaches to employing a Deep CNN (DCNN) depending on the type of problem. In this regard, one of the existing methods is to use pre-trained models. A pre-trained model is a model created and trained on a large dataset to solve a problem. These models are available as open-source.

In this research, image features were extracted by using pre-trained networks in the first step. Then, those features were given to the ANOVA algorithm to reduce features and select a number of them to improve the performance. Then the selected features were provided as input to the LightGBM to classify the images. The remainder of the paper is structured as follows: In section II, related researches will be reviewed. Section III provides the required background topics like the LightGBM method. In section IV, the proposed approach will be introduced. The results will be reported and discussed in section V. Finally, in section VI, a summary of the findings and conclusion will be presented.

## II. RELATED WORKS

Wang et al. [16] developed a deep convolutional neural network called COVID-Net by using CXR images, which helped physicians in the screening phase. In their research, Narin et al. [17] presented the five previously trained CNN models of ResNet50, ResNet101, ResNet152, InceptionV3, and Inception-ResNetV2 to diagnose patients with pneumonia and COVID-19 using their patients' CXR images. Abbas et al. [18] developed a DCNN called DeTraC to detect COVID-19. In their network, diagnosis and classification were done using 80 images of healthy people and 126 images of people with Corona disease. Their proposed method was able to reach 95.12% accuracy. Hemdan et al. [19] studied 50 CXR images, including 25 cases of COVID-19 infection. In their study, a model called COVIDX-Net was presented. Seven different architectures of DCNN models, such as VGG19 and Google MobileNet, were employed in this model. Each DCNN model was able to analyze images and classify healthy individuals vs. patients with COVID-19. In their research, Hemdan et al. expressed the best performances for VGG19 and DenseNet models with 89% and 91% accuracies, respectively. Tabik et al. [20] used a dataset called COVIDGR-1.0, which contained 426 images of those with COVID-19 and 426 images of healthy people. In their proposed method called COVID-SDNet, they tried to improve the diagnosis and classification of healthy people and patients with COVID-19 at different levels. They achieved accuracies of 97.72%, 86.90%, and 67.80% for severe, moderate, and weak conditions of a person with COVID-19, respectively.

Minaee et al. [21] applied a dataset of 5,000 CXR images of patients with the 4 CNNs of ResNet50, ResNet18, SqueezeNet, and DenseNet-121 to diagnose COVID-19 disease. In their investigation, Iqbal Khan et al. [22] proposed a CoroNet model, which was based on a pre-trained Xception architecture, by using a DCNN for automatic diagnosis of COVID-19 disease. In the proposed CoroNet model, they achieved 95% accuracy for the multi-class problem (i.e., COVID-19, Pneumonia, and No-findings). Ozturk et al. [11] proposed a Deep Neural Network (DNN) based on the DarkNet model. In their study, they utilized 1,125 CT-scan images, including 125, 500, and 500 images of patients' chests with Corona disease, pneumonia, and no-findings, respectively. In their two-class and multi-class classification problems, they reached accuracies of 98.08% and 87.02%, respectively. Nasiri and Hasani [23] employed a DNN, i.e., DenseNet169, and the XGBoost algorithm to extract image features and classify them, respectively. They attained average accuracies of 98.24% and 89.70% in two-class and multi-class problems, respectively.

III. BACKGROUNDS

This section introduces the DenseNet169, LightGBM algorithm, and Analysis of Variance (ANOVA) employed in the proposed method.

*A. DenseNet169*

DenseNet169 is one of the DenseNet models designed for image classification. All the DenseNet models are pre-trained on the ImageNet dataset. Each layer in the DenseNet receives inputs from the previous layers and sends its output to the next layers. Since each layer receives features from all the previous layers, the network can become more compact, resulting in fewer channels, which improves computational performance. Its main difference with the DenseNet121 model is in its size and accuracy. The DenseNet169 is about 55 MB larger than the DenseNet121 (315 MB). This network has been able to achieve 93.20% accuracy in the ImageNet dataset [24].

*B. LightGBM*

The Light Gradient Boosting Machine (LightGBM) algorithm is an open-source, distributed, high-performance Gradient-Boosting (GB) framework, which has been originally introduced by Microsoft [25], [26]. Ke et al. [25] proposed a decision tree for efficient GB using a Gradient-based One-way Sampling Solution (GOSS) and an Exclusive Feature Bundling (EFB) called LightGBM [27]. Unlike other boosting algorithms that grow a tree level by level, LightGBM grows it with the help of GOSS and a leaf-wise tree growth approach. The two GOSS and EFB techniques discussed below form the basis of the LightGBM algorithm [25], [27].

**Gradient-based One-way Sampling Solution (GOSS)**: Samples with different gradients play different roles in calculating information gain. According to the information gain definition, samples with larger gradients have a greater role in the information gain. As a result, GOSS retains samples with larger gradients and randomly drops those with smaller ones. This provides higher estimation accuracy compared to a uniform random sampling with the same sampling rate. Mathematically speaking, consider a training dataset with n samples $\{\mathbf{x}_1, \mathbf{x}_2, \cdots, \mathbf{x}_n\}$, where $\mathbf{x}_i$ is a vector with dimension $s$ in the input space $\chi^S$. In each iteration of gradient boosting, the negative gradient of the loss function relative to the model's output is shown as $\{g_1, g_2, \cdots, g_n\}$. In the GOSS method, the training samples are first sorted in a descending order according to the absolute values of their gradients. Then, top $a\%$ samples with larger gradients are stored in a subset like $A$. Afterwards, for the remaining set $A^C$, which contains $(1-a)\%$ of the samples with smaller gradients, a subset $B$ with the size of $b \times |A^C|$ is randomly sampled. Finally, the samples are split over the set $A \cup B$ according to the estimated variance gain $\tilde{V}_j(d)$, which can be calculated using (1) [25], [27], [28].

$$\tilde{V}_j(d) = \frac{1}{n}\left( \frac{(\sum_{x_i \in A_l} g_i + \frac{1-a}{b}\sum_{x_i \in B_l} g_i)^2}{n_l^j(d)} + \frac{(\sum_{x_i \in A_r} g_i + \frac{1-a}{b}\sum_{x_i \in B_r} g_i)^2}{n_r^j(d)} \right) \quad (1)$$

where $A_l = \{x_i \in A : x_{ij} \leq d\}$, $A_r = \{x_i \in A : x_{ij} > d\}$, $B_l = \{x_i \in B : x_{ij} \leq d\}$, $B_r = \{x_i \in B : x_{ij} > d\}$, and $\frac{1-a}{b}$ is utilized to normalize the gradients [25], [27].

**Exclusive Feature Bundling (EFB):** Data in a high-dimensional space is usually very sparse and analyzing it has many challenges [29]. The sparsity of the feature space makes it possible to reduce the number of features by designing a lossless method. In the EFB method, several features are bundled into a single one. Thus, the computational complexity of LightGBM changes from $O(\#data \times \#feature)$ to $O(\#data \times \#bundle)$ and since $\#bundle \ll \#features$, the training speed increases without decreasing accuracy [25], [27], [28].

In general, the LightGBM algorithm is an improved and much more efficient version of the Gradient-Boosted Decision Tree (GBDT) method, like XGBoost [27], [30]–[32]. This algorithm has been proven to speed up the conventional GBDT training process by more than 20 times while achieving almost the same accuracy. The output predicted by the LightGBM model is given in the following formula:

$$F_M(\mathbf{x}) = \sum_{m=1}^{M} h_m(\mathbf{x}) \qquad (2)$$

where $M$ and $h_m(\mathbf{x})$ denote the maximum number of iterations, and the base decision tree, respectively.

*C. ANOVA Feature Selection Method*

Feature selection is the process of identifying the relevant features and removing the irrelevant and redundant ones to reduce the number of input variables when developing a model. By decreasing the input variables, the model's computational cost is reduced and its accuracy is enhanced in some cases. The ANOVA method is a popular statistical approach for comparing several independent means. In this method, the features are scored based on the ratio of the between-group variance to the within-group variance through the F-test [33]. The statistical ratio of the F-test is given in (3).

$$F = \frac{\text{Variance between samples}}{\text{Variance within samples}} \qquad (3)$$

## IV. PROPOSED METHOD

In this research, the pre-trained DenseNet169 model was first used to extract image features. Then, the last layer of the DenseNet169, which was configured to predict ImageNet dataset classes, was removed. To reduce the number of extraction features that could lower the computational complexity and avoid the risk of overfitting, an average pooling layer was added to the end of the network. Then, the extracted features were given to the ANOVA features selection algorithm to reduce features and select a number of them to improve the performance. Finally, the selected features were provided as input to the LightGBM algorithm to classify the images and detect COVID-19 cases. The LightGBM was trained by using the feature vector of the input images and their corresponding labels. The proposed method's framework is depicted in Fig. 2.

## V. EXPERIMENTAL RESULTS

The ChestX-ray8 dataset [34] was employed to evaluate the proposed algorithm. This dataset included a total of 1125 images, consisting of 125, 500, and 500 images of patients' chests with Corona disease, pneumonia, and no-findings, respectively. Fig. 1 shows the sample images in the dataset. For evaluating the proposed method, the experiments were carried out in two ways, while in both experiments, 60%, 20%, and 20% of the data were utilized for train, validation, and test, respectively. In the first experiment, the multi-class problem (i.e., COVID-19, Pneumonia, and No-findings), 1125 images, and in the second experiment, the two-class problem (i.e., COVID-19, No-findings), 625 images were taken into account.

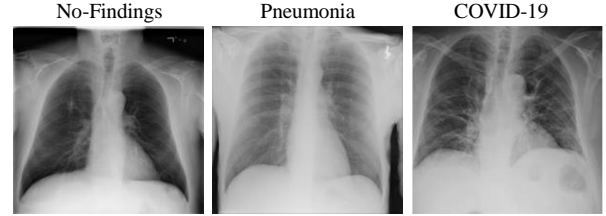

Fig. 1. Sample Images in the Dataset

The best trained DNN was chosen among eight networks in the first stage. Table I shows the results of this experiment. As the results show, the DenseNet169 network has the highest accuracy and has been selected as the desired network to be applied in the proposed method. The output of this network is a feature vector with 1664 features. In the second step, the ANOVA algorithm was utilized to select the features. According to the results obtained from the validation set, 116 and 133 features were selected in the multi-class and two-class classification problems, respectively. Finally, in the last step, the parameters of the LightGBM algorithm were tuned by trial and error. The mentioned parameters are presented in Table II.

TABLE I. THE AVERAGE ACCURACY COMPARISON OF DIFFERENT DNNs

| DNN | Average Accuracy (%) | |
| --- | --- | --- |
|  | Multi-Class Problem | Two-Class Problem |
| Xception | 75.11 | 94.00 |
| ResNet50 | 77.33 | 96.40 |
| ResNet101 | 80.89 | 97.40 |
| ResNet152 | 79.86 | 94.40 |
| InceptionV3 | 76.00 | 93.60 |
| MobileNet | 78.67 | 96.40 |
| DenseNet121 | 82.67 | 97.00 |
| **DenseNet169** | **82.67** | **97.40** |

TABLE II. THE LIGHTGBM PARAMETER SETTINGS

| Parameter | Multi-Class Problem | Two-Class Problem |
| --- | --- | --- |
| Base Learner | Gradient boosted tree | |
| Tree construction algorithm | Exact greedy | |
| Number of gradients boosted trees | 200 | 100 |
| Learning rate $(\eta)$ | 0.24 | 0.20 |
| Lagrange multiplier $(\gamma)$ | 0 | 0 |
| Maximum tree depth | 3 | 0 |

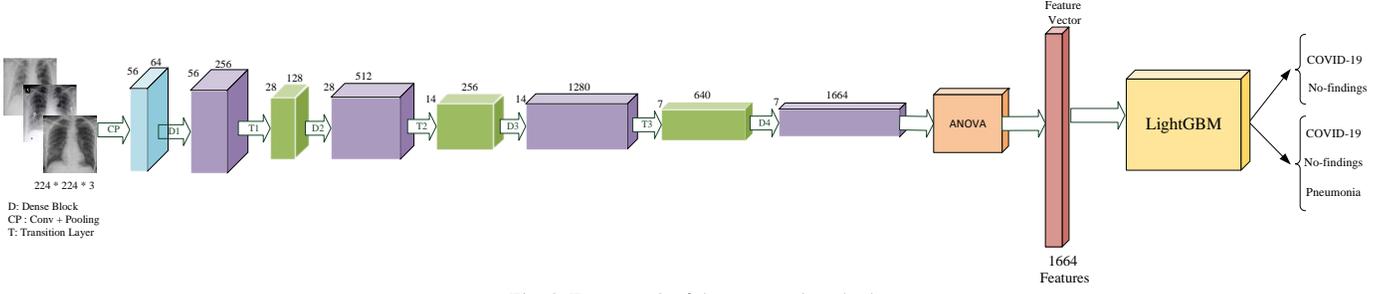

Fig. 2. Framework of the proposed method

After tuning the parameters, the accuracy of the proposed method was obtained by using the test set. The accuracy for multi-class problem was 94.22% and the average accuracy for two-class problem was 99.20%. It should be noted that in the two-class problem, the 5-fold cross-validation method was employed. The confusion matrix for the multi-class problem can be seen in Fig. 3. Also, the confusion matrix for each fold of the two-class problem and all the data set is depicted in Fig. 4.

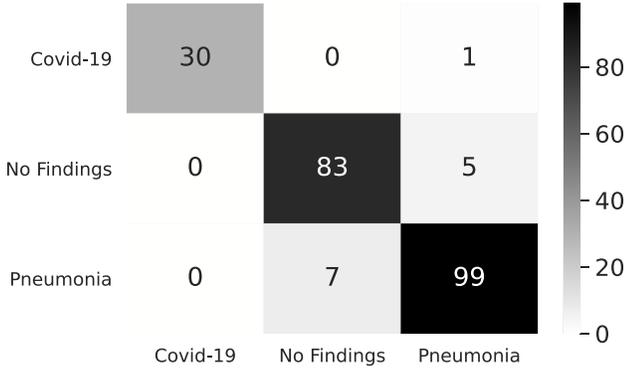

Fig. 3. The Confusion Matrix (Multi-Class Problem)

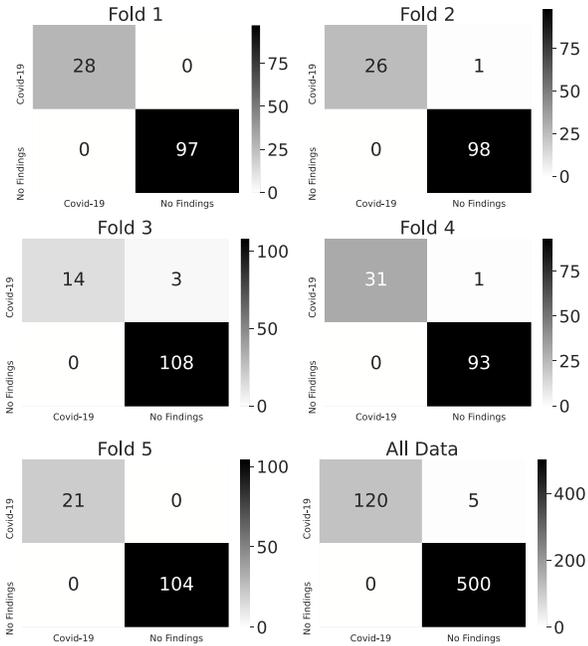

Fig. 4. The Confusion Matrix (Two-Class Problem)

To compare the proposed method with other methods, the following evaluation criteria were applied:

**Sensitivity:** Mathematically, it is the result of dividing true positive cases by the sum of true positive and false negative cases as shown in (4).

$$Sensitivity = \frac{TP}{TP + FN} \quad (4)$$

**Specificity:** Mathematically, it is the result of dividing true negative cases by the sum of true negative and false positive cases as shown in (5):

$$Specificity = \frac{TN}{TN + FP} \quad (5)$$

**Precision:** Mathematically, it is the result of dividing true positive cases by the sum of true positive and false positive cases as shown in (6).

$$Precision = \frac{TP}{TP + FP} \quad (6)$$

**Accuracy:** Mathematically, it is the result of dividing the sum of true negative and true positive cases by the sum of true negative, true positive, false negative, and false positive cases (total number of cases) as shown in (7).

$$Accuracy = \frac{TN + TP}{TN + TP + FN + FP} \quad (7)$$

**F$_1$-Score:** Mathematically speaking, it is the result of dividing the doubled precision and sensitivity by their sum as shown in (8) [33].

$$F_1\text{-}Score = \frac{2 \times (Precision \times Sensitivity)}{Precision + Sensitivity} \quad (8)$$

TABLE III. THE COMPARISON OF THE PROPOSED METHOD WITH OTHER METHODS (MULTI-CLASS PROBLEM)

|  | Proposed Method | Nasiri and Hasani | Ozturk et al. |
|---|---|---|---|
| Sensitivity | **100** | 95.20 | 85.35 |
| Specificity | **100** | **100** | 92.18 |
| Precision | **95.50** | 92.50 | 89.96 |
| F1-score | **95.15** | 91.20 | 87.37 |
| Accuracy | **94.22** | 89.70 | 87.02 |

Tables III and Table IV represent the comparison of the proposed method with other methods in two-class and multi-class problems, respectively. As can be seen from the results of these tables, the proposed method has performed better than the other methods in both problems. Table V compares the proposed method with other DL methods, in which chest images have been employed. As can be observed, the proposed method in this research has better accuracy than other proposed methods. Of course, the results shown in Table V have not been tested on the same dataset.

TABLE IV. COMPARISON OF THE PROPOSED METHOD WITH OTHER METHODS (TWO-CLASS PROBLEM)

| Performance Metrics | | Fold 1 | Fold 2 | Fold 3 | Fold 4 | Fold 5 | Average |
|---|---|---|---|---|---|---|---|
| Sensitivity | Proposed Method | **100** | **100** | **100** | **100** | **100** | **100** |
| | Nasiri and Hasani | 95.20 | 95.40 | 96.70 | 81.40 | 91.40 | 92.08 |
| | Ozturk et al. | **100** | 96.42 | 90.47 | 93.75 | 93.18 | 95.13 |
| Specificity | Proposed Method | **100** | **99.20** | 82.25 | **96.87** | **100** | 95.10 |
| | Nasiri and Hasani | **100** | **100** | **100** | 89.90 | **100** | **99.78** |
| | Ozturk et al. | 100 | 96.42 | 90.47 | 93.75 | 93.18 | 95.30 |
| Precision | Proposed Method | **100** | 98.99 | 97.30 | **98.94** | **100** | **99.04** |
| | Nasiri and Hasani | 99.50 | **99.50** | **99.40** | 95.30 | 99.02 | 98.54 |
| | Ozturk et al. | **100** | 94.52 | 98.14 | 98.57 | 98.58 | 98.03 |
| F1-score | Proposed Method | **100** | **98.80** | 94.48 | **98.94** | **100** | **98.44** |
| | Nasiri and Hasani | 98.50 | 98.50 | **98.20** | 92.50 | 97.30 | 97.00 |
| | Ozturk et al. | **100** | 95.52 | 93.79 | 95.93 | 95.42 | 96.51 |
| Accuracy | Proposed Method | **100** | **99.20** | 97.60 | **99.20** | **100** | **99.20** |
| | Nasiri and Hasani | 99.20 | **99.20** | **99.20** | 95.20 | 98.40 | 98.24 |
| | Ozturk et al. | **100** | 97.60 | 96.80 | 97.60 | 97.60 | 98.08 |

TABLE V. COMPARISON OF THE PROPOSED METHOD WITH OTHER DL-BASED METHODS

| Study | Type of Images | Number of Samples | Method Used | Accuracy (%) |
|---|---|---|---|---|
| Apostolopoulos et al. [35] | Chest X-ray | 1428 | VGG-19 | 93.48 |
| Wang et al. [16] | Chest X-ray | 13645 | COVID-Net | 92.40 |
| Sethy et al. [36] | Chest X-ray | 50 | ResNet50 + SVM | 95.38 |
| Hemdan et al. [19] | Chest X-ray | 50 | COVIDX-Net | 90.00 |
| Narin et al. [17] | Chest X-ray | 100 | Deep CNN ResNet-50 | 98.00 |
| Song et al. [37] | Chest CT | 1485 | DRE-Net | 86.00 |
| Wang et al. [38] | Chest CT | 453 | M-Inception | 82.90 |
| Zheng et al. [39] | Chest CT | 542 | UNet + 3D Deep Network | 90.80 |
| Xu et al. [40] | Chest CT | 443 | ResNet + Location Attention | 86.60 |
| Ozturk et al. [11] | Chest X-ray | 625 | DarkCovidNet | 98.08 |
| | | 1125 | | 87.02 |
| Nasiri and Hasani [23] | Chest X-ray | 625 | DenseNet169 + XGBoost | 98.23 |
| | | 1125 | | 89.70 |
| Proposed Method | Chest X-ray | 625 | DenseNet169 + LightGBM | **99.20** |
| | | 1125 | | **94.22** |

## VI. Conclusion

In this paper, using artificial intelligence techniques, a model was proposed for diagnosing people with COVID-19. The DenseNet169 was employed in the proposed approach to extract image features. The number of the extracted features was reduced by using the ANOVA feature selection algorithm and finally, the LightGBM algorithm performed image classification to diagnose COVID-19. The results obtained from the evaluation experiments revealed that the proposed method had the accuracies of 94.22% and 99.20% in the multi-class and two-class problems, respectively, which still showed higher accuracies than those of the other methods. It should be noted that unlike the methods presented by other researchers, only the LightGBM algorithm was trained in the proposed method, while the DNN did not need any training, thus enhancing the speed of this method and reducing its complexity.

## References

[1] F. Wu *et al.*, "A new coronavirus associated with human respiratory disease in China," *Nature*, vol. 579, no. 7798, pp. 265–269, 2020, doi: 10.1038/s41586-020-2008-3.

[2] K. Roosa *et al.*, "Real-time forecasts of the COVID-19 epidemic in China from February 5th to February 24th, 2020," *Infect. Dis. Model.*, vol. 5, pp. 256–263, Jan. 2020, doi: 10.1016/J.IDM.2020.02.002.

[3] S. B. Stoecklin *et al.*, "First cases of coronavirus disease 2019 (COVID-19) in France: surveillance, investigations and control measures, January 2020," *Eurosurveillance*, vol. 25, no. 6, p. 2000094, Feb. 2020, doi: 10.2807/1560-7917.ES.2020.25.6.2000094.

[4] T. Singhal, "A review of coronavirus disease-2019 (COVID-19)," *indian J. Pediatr.*, vol. 87, no. 4, pp. 281–286, 2020.

[5] S. Hasani and H. Nasiri, "COV-ADSX: An Automated Detection System using X-ray images, deep learning, and XGBoost for COVID-19," *Softw. Impacts*, vol. 11, p. 100210, 2022, doi: 10.1016/j.simpa.2021.100210.

[6] L. Cirrincione *et al.*, "COVID-19 pandemic: Prevention and protection measures to be adopted at the workplace," *Sustainability*, vol. 12, no. 9, p. 3603, 2020.

[7] Z. Y. Zu *et al.*, "Coronavirus disease 2019 (COVID-19): a perspective from China," *Radiology*, vol. 296, no. 2, pp. E15--E25, 2020.

[8] P. Huang *et al.*, "Use of chest CT in combination with negative RT-PCR assay for the 2019 novel coronavirus but high clinical suspicion," *Radiology*, vol. 295, no. 1, pp. 22–23, 2020.

[9] R. G. Babukarthik, V. A. K. Adiga, G. Sambasivam, D. Chandramohan, and J. Amudhavel, "Prediction of covid-19 using genetic deep learning convolutional neural network (GDCNN)," *IEEE Access*, vol. 8, pp. 177647–177666, 2020.

[10] H. Y. F. Wong *et al.*, "Frequency and distribution of chest radiographic findings in patients positive for COVID-19," *Radiology*, vol. 296, no. 2, pp. E72--E78, 2020.

[11] T. Ozturk, M. Talo, E. A. Yildirim, U. B. Baloglu, O. Yildirim, and U. R. Acharya, "Automated detection of COVID-19 cases using deep neural networks with X-ray images," *Comput. Biol. Med.*, vol. 121, p. 103792, 2020.

[12] R. Yamashita, M. Nishio, R. K. G. Do, and K. Togashi, "Convolutional neural networks: an overview and application in radiology," *Insights into Imaging 2018 94*, vol. 9, no. 4, pp. 611–629, Jun. 2018, doi: 10.1007/S13244-018-0639-9.

[13] L. Alzubaidi *et al.*, *Review of deep learning: concepts, CNN architectures, challenges, applications, future directions*, vol. 8, no. 1. Springer International Publishing, 2021.

[14] L. Sha, "Efficient Methods in Deep Learning Lifecycle: Representation, Prediction and Model Compression," Brandeis University, 2021.

[15] S. Indolia, A. K. Goswami, S. P. Mishra, and P. Asopa, "Conceptual Understanding of Convolutional Neural Network- A Deep Learning Approach," *Procedia Comput. Sci.*, vol. 132, pp. 679–688, Jan. 2018, doi: 10.1016/J.PROCS.2018.05.069.

[16] L. Wang, Z. Q. Lin, and A. Wong, "Covid-net: A tailored deep convolutional neural network design for detection of covid-19 cases from chest x-ray images," *Sci. Rep.*, vol. 10, no. 1, pp. 1–12, 2020.

[17] A. Narin, C. Kaya, and Z. Pamuk, "Automatic detection of coronavirus disease (covid-19) using x-ray images and deep convolutional neural networks," *Pattern Anal. Appl.*, pp. 1–14, 2021.

[18] S. Asif, Y. Wenhui, H. Jin, and S. Jinhai, "Classification of COVID-19 from Chest X-ray images using Deep Convolutional Neural Network," *2020 IEEE 6th Int. Conf. Comput. Commun. ICCC 2020*, pp. 426–433, 2020, doi: 10.1109/ICCC51575.2020.9344870.

[19] E. E.-D. Hemdan, M. A. Shouman, and M. E. Karar, "Covidx-net: A framework of deep learning classifiers to diagnose covid-19 in x-ray images," *arXiv Prepr. arXiv:2003.11055*, 2020.

[20] S. Tabik *et al.*, "COVIDGR Dataset and COVID-SDNet Methodology for Predicting COVID-19 Based on Chest X-Ray Images," *IEEE J. Biomed. Heal. Informatics*, vol. 24, no. 12, pp. 3595–3605, 2020, doi: 10.1109/JBHI.2020.3037127.

[21] S. Minaee, R. Kafieh, M. Sonka, S. Yazdani, and G. J. Soufi, "Deep-COVID: Predicting COVID-19 from chest X-ray images using deep transfer learning," *Med. Image Anal.*, vol. 65, 2020, doi: 10.1016/j.media.2020.101794.

[22] A. I. Khan, J. L. Shah, and M. M. Bhat, "CoroNet: A deep neural network for detection and diagnosis of COVID-19 from chest x-ray images," *Comput. Methods Programs Biomed.*, vol. 196, p. 105581,

2020.

[23] H. Nasiri and S. Hasani, "Automated detection of COVID-19 cases from chest X-ray images using deep neural network and XGBoost," *arXiv Prepr. arXiv:2109.02428*, 2021, [Online]. Available: http://arxiv.org/abs/2109.02428.

[24] G. Huang, Z. Liu, L. Van Der Maaten, and K. Q. Weinberger, "Densely connected convolutional networks," in *Proceedings of the IEEE conference on computer vision and pattern recognition*, 2017, pp. 4700–4708.

[25] G. Ke *et al.*, "Lightgbm: A highly efficient gradient boosting decision tree," *Adv. Neural Inf. Process. Syst.*, vol. 30, pp. 3146–3154, 2017.

[26] Q. Meng *et al.*, "A communication-efficient parallel algorithm for decision tree," *arXiv Prepr. arXiv:1611.01276*, 2016.

[27] C. Chen, Q. Zhang, Q. Ma, and B. Yu, "LightGBM-PPI: Predicting protein-protein interactions through LightGBM with multi-information fusion," *Chemom. Intell. Lab. Syst.*, vol. 191, pp. 54–64, 2019.

[28] M. R. Machado, S. Karray, and I. T. de Sousa, "LightGBM: An effective decision tree gradient boosting method to predict customer loyalty in the finance industry," in *2019 14th International Conference on Computer Science \& Education (ICCSE)*, 2019, pp. 1111–1116.

[29] A. Hemmati, H. Nasiri, M. A. Haeri, and M. M. Ebadzadeh, "A Novel Correlation-Based CUR Matrix Decomposition Method," in *2020 6th International Conference on Web Research (ICWR)*, 2020, pp. 172–176.

[30] S. Chehreh Chelgani, H. Nasiri, and A. Tohry, "Modeling of particle sizes for industrial HPGR products by a unique explainable AI tool- A 'Conscious Lab' development," *Adv. Powder Technol.*, vol. 32, no. 11, pp. 4141–4148, 2021, doi: 10.1016/j.apt.2021.09.020.

[31] S. C. Chelgani, H. Nasiri, and M. Alidokht, "Interpretable modeling of metallurgical responses for an industrial coal column flotation circuit by XGBoost and SHAP-A 'conscious-lab' development," *Int. J. Min. Sci. Technol.*, vol. 31, no. 6, pp. 1135–1144, 2021, doi: 10.1016/j.ijmst.2021.10.006.

[32] H. Nasiri, A. Homafar, and S. Chehreh Chelgani, "Prediction of uniaxial compressive strength and modulus of elasticity for Travertine samples using an explainable artificial intelligence," *Results Geophys. Sci.*, vol. 8, p. 100034, 2021, doi: 10.1016/j.ringps.2021.100034.

[33] H. Nasiri and S. A. Alavi, "A Novel Framework Based on Deep Learning and ANOVA Feature Selection Method for Diagnosis of COVID-19 Cases from Chest X-Ray Images," *Comput. Intell. Neurosci.*, vol. 2022, p. 4694567, 2022, doi: 10.1155/2022/4694567.

[34] X. Wang, Y. Peng, L. Lu, Z. Lu, M. Bagheri, and R. M. Summers, "ChestX-ray: Hospital-Scale Chest X-ray Database and Benchmarks on Weakly Supervised Classification and Localization of Common Thorax Diseases," *Adv. Comput. Vis. Pattern Recognit.*, pp. 369–392, 2019, doi: 10.1007/978-3-030-13969-8_18.

[35] I. D. Apostolopoulos and T. A. Mpesiana, "Covid-19: automatic detection from x-ray images utilizing transfer learning with convolutional neural networks," *Phys. Eng. Sci. Med.*, vol. 43, no. 2, pp. 635–640, 2020.

[36] P. K. Sethy, S. K. Behera, P. K. Ratha, and P. Biswas, "Detection of coronavirus disease (COVID-19) based on deep features and support vector machine," *Int. J. Math. Eng. Manag. Sci.*, vol. 5, no. 4, pp. 643–651, 2020.

[37] Y. Song *et al.*, "Deep learning enables accurate diagnosis of novel coronavirus (COVID-19) with CT images," *IEEE/ACM Trans. Comput. Biol. Bioinforma.*, vol. 18, no. 6, pp. 2775–2780, 2021.

[38] S. Wang *et al.*, "A deep learning algorithm using CT images to screen for Corona Virus Disease (COVID-19)," *Eur. Radiol.*, pp. 1–9, 2021.

[39] C. Zheng *et al.*, "Deep learning-based detection for COVID-19 from chest CT using weak label," *MedRxiv*, 2020.

[40] X. Xu *et al.*, "A deep learning system to screen novel coronavirus disease 2019 pneumonia," *Engineering*, vol. 6, no. 10, pp. 1122–1129, 2020.